# Wide band, tunable gamma-ray lenses

Niels Lund

DTU Space, Technical University of Denmark
Lundtoftevej 327, DK 2800, Kgs. Lyngby, Denmark
nl@dtu.space.dk

**Abstract.** A new concept for an astronomical telescope in the MeV energy band is presented. The concept builds on Bragg diffraction in crystals, which has been discussed in the past, but so far a design with good sensitivity over a wide energy range has seemed out of reach. In this paper we point out that if we find ways to adjust, in orbit, the individual tilt of all the crystals in the lens this would allow one single lens to cover with excellent efficiency the full range of energies from 200 keV to 2.5 MeV in a few observation steps. Secondly, we note that the use of lenses with double crystal layers will increase the photon collection significantly. In an accompanying paper we describe our overall lens design in more detail and present our first prototype tilt adjustment pedestal for use with the individual lens facets.

## 1) Introduction

From radio to X-rays focusing telescopes are at the forefront of the astronomical research. Unfortunately, focusing techniques based on normal incidence mirrors are not feasible at energies beyond a few tens of eV, and grazing incidence mirrors, even with multilayer coatings becomes impractical at energies beyond a few hundred keV. This is just where we enter the energy range of nuclear gamma rays and electron-positron annihilation astrophysics. So a focusing telescope – or just a non-imaging, flux collection telescope – in the MeV energy range would be a most welcome addition to our observational capabilities.

Among the scientific problems that awaits a major improvement in the sensitivity of our MeV instrumentation we would like to emphasize a few which would benefit from Laue lens technology:

a) Identify the sites in which the majority of the r-process elements in the Universe have been formed. The origin of the r-process elements is currently a hot topic in the astrophysical community and direct detection of nuclear line emission associated with specific types of explosive stellar events could resolve this uncertainty (Lattimer and Schramm 1976; Arnould et al. 2007; Sneden et al. 2008; Hotokezaka et al. 2015; Siegel et al. 2019; Watson et al. 2019).
b) The detection of nuclear lines related to X-ray bursts or nova eruptions would impact the models used to describe these events. Similarly, the possible detection of emission from electron annihilation or deuterium formation in the accretion disk in connection with bursts and eruptions would help to provide a more complete picture of the accretion process.
c) Detection of 511 keV emission from the centers of nearby normal galaxies and more distant AGNs would put the 511 keV emission from our own Galactic Center in perspective.(Kierans et al. 2019)

The idea to use crystals in the Laue-configuration for focusing hard X-rays is not new. The first attempts to build such ‘Laue-lenses’ goes back to the 1960’s (Lindquist and Webber 1968). A pioneering balloon flight demonstrating the Laue lens concept was carried out by the Toulouse group in 2001 (Halloin et al. 2003), and a demonstration experiment probing the feasibility of crystal tilt adjustment was reported by the same group some years before (Kohnle et al. 1998). However, in both cases, the technologies used for the crystal mounting was not suited for up-scaling to a large satellite experiment.

## 2) The classic Laue lens

The classic idea of a Laue-lens is to arrange a number of crystals in a circular pattern assuring that the diffracted beams from all crystals intercept at a common focus. At the core of this concept is Bragg's diffraction law:

$$\sin(\theta) = N^*\lambda / 2d \qquad \text{with } N = 1,2,3,\ldots$$

where θ is the incidence angle of the photon with respect to the selected crystal plane, $\lambda$ is the photon wavelength ($\lambda \infty 1/E$) and $d$ is the crystal plane separation.

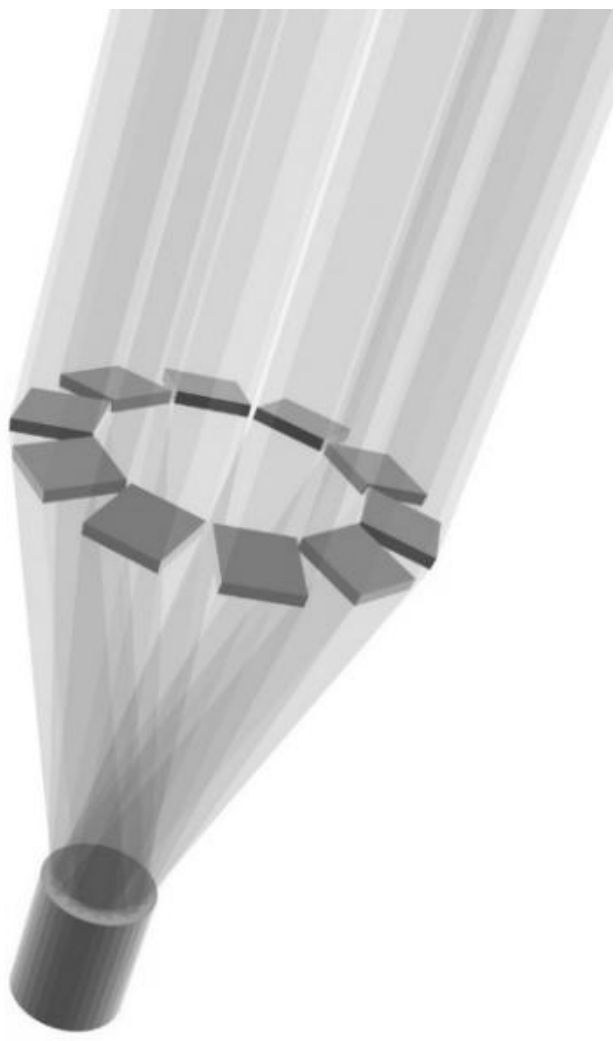

**Fig 1** Laue lens concept

Bragg's relation is unforgiving: diffraction will only take place if a photon with the right wavelength traverses a crystal with the right *d*-spacing set at the correct tilt angle, θ. Therefore, each crystal can only diffract a narrow band of energies, and this constrains significantly the overall efficiency of Laue lenses. Through use of crystals with some internal disorder, 'mosaicity', one can achieve diffraction over a limited angular (and energy-) range, but this comes at a cost, because it implies a corresponding angular divergence of the diffracted photons, and consequently some smearing of the focal spot.

The simplicity of the classic Laue lens is deceptive. Firstly, the crystals used must be of uniform, high quality, meaning that their mosaic structure should not exhibit distinct regions misaligned with respect to each other. The technology of growing such crystals in large dimension have already been demonstrated. (Courtois et al., 2005). Secondly, a large number (several thousand) crystals are required to construct a lens with a significant gain over an extended energy range. And thirdly, the crystals must be mounted with high precision (±5 arc seconds in our case) on a large and stable platform (3.5 m in diameter for the lens discussed here). The requirements on the crystal mounting has turned out to be very difficult to fulfill. The primary obstacle apparently is that most adhesives typically exhibits shrinkage during the setting and hardening process so the crystals twists and maybe even deforms during this process. Mosaic Laue crystals are quite soft and fragile. Moreover, it must be foreseen that not only the crystals, but also the support platform deforms under the glue stresses. The Laue lens platform is not just a support for a thin reflection coating as is the case for the mirror blank of an optical telescope, but the Laue platform must carry a substantial mass of crystals and must still be transparent to the gamma rays, thus not a very massive structure. To our knowledge this point has not been addressed seriously by any Laue lens study performed up to now.

3) **A new lens configuration.**

The practical problems associated with the crystal mounting prompted us to look for a new way to set up a Laue lens. We propose to use an active optical alignment scheme for the individual crystals in the lens as this will significantly relax the requirements on the platform stability. Of course the active control is a major complication, but as we shall show in the accompanying technology paper (Lund, 2020), not an insurmountable one. Moreover, this scheme bring us the added possibility to retune our lens for different energy ranges, which opens up important new astrophysical possibilities.

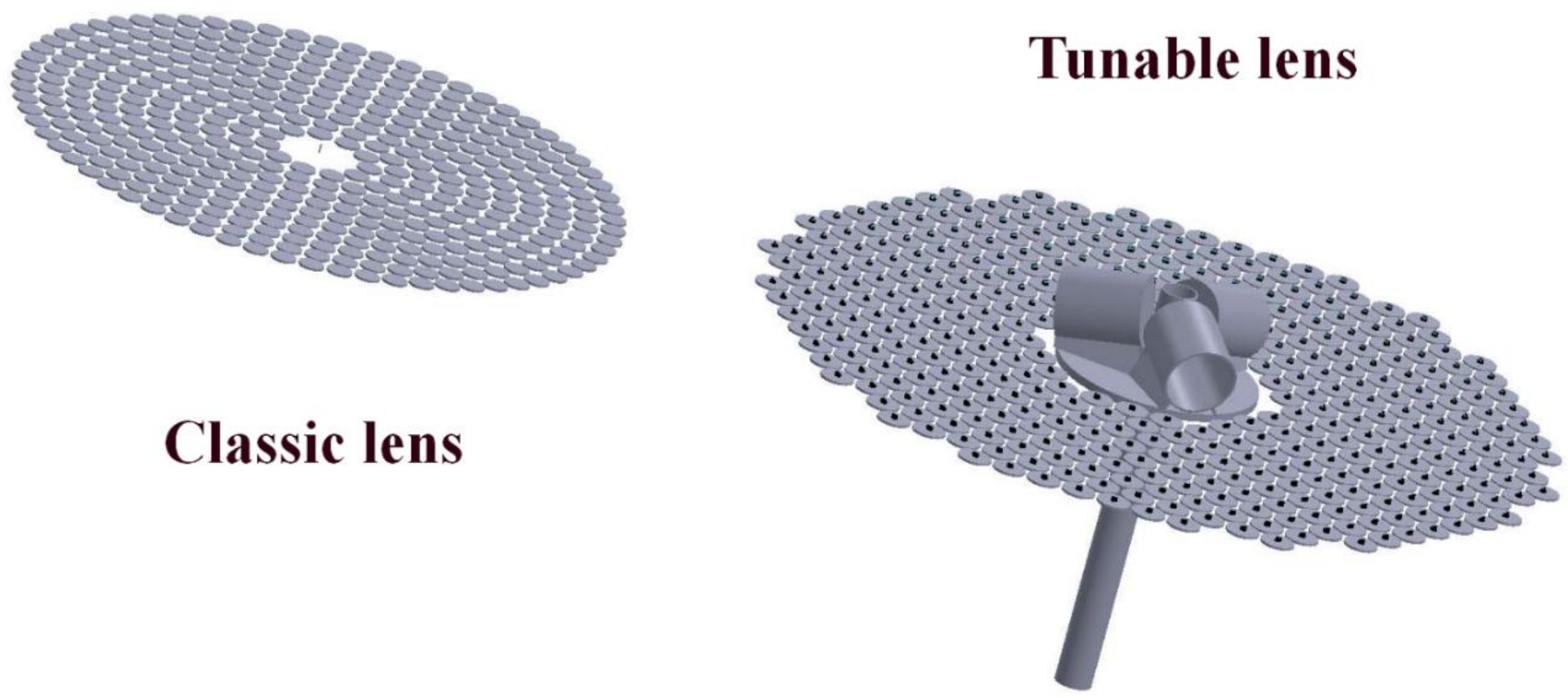

**Fig. 2** Comparison of the classic, fixed energy Laue lens and the tunable lens. In the center of the tunable lens are located three optical autocollimator telescopes on a scanning platform. These telescopes are the key elements for the in-flight tuning and retuning of the crystal tilt angles. In the center of each crystal is a small alignment mirror visible from the autocollimators. Drawing not to scale.

The use of an integrated optical alignment system is central to our concept. This system is indicated by the three cylinders (autocollimator telescopes) shown on the central circular platform in Figure 2. Each crystal in our lens is mounted on an adjustable pedestal and equipped with an alignment mirror. The central platform can rotate and scan all the crystals and it is the axis of this rotating platform which defines the gamma-ray optical axis of the Laue lens. The orientation of the alignment mirror with respect to the Bragg plane of its crystal is determined with arc second precision early in our crystal preparation process. All later alignment operations relies on the optical mirrors.

We have based our lens on the use of conventional, mosaic crystals of silver and copper. We plan to employ rather large crystals (40 mm diameter) and in this paper we assume a detector size of 50 mm diameter. We accept that our telescope will not be an imaging instrument, but just a flux collector. We have modelled our lens on a circular platform, 3.5 m in diameter covered with circular crystals in hexagonal close packing with 2.5 mm separation of the crystals. A hexagonal arrangement of circular crystals may seem inefficient in terms of utilization of the lens area, but it is convenient for our proposed tilt-adjustable lens pedestals and for the double crystal layers both of which are key elements in our concept.

We have refrained from the use of 'advanced' (Smither et al. 2006; Camattari et al. 2018) or 'designed' (Ackermann et al. 2013) Laue-crystals in our simulations. These rely on 'perfect' crystals, in practice: Silicon or Germanium. However, for the high energy application we have in mind Silicon is not well suited due to its small atomic form factor and even the improvement (by a factor 2) promised by the 'advanced' designs still does not make Silicon an attractive candidate for our purposes. Germanium remains an interesting possibility, but the production of bent crystal plane Germanium crystals of the dimensions required for this application is still unproven.

We use Darwin's model and the dynamic theory of X-ray diffraction throughout in the paper. The optimal thickness for a Laue crystal is found by maximizing the product of the diffraction probability (rising with crystal thickness) and the transmission probability (falling with crystal thickness). The product of the two probabilities has a flat maximum as function of the crystal thickness. Further details about Laue telescopes can be found in the literature (Lund 1992; Halloin and Bastie 2005).

### 4) Performance of the classic Laue-lens

A lens constructed from only one type of crystal, all set to use the same crystal planes for diffraction will have an effective collection area which varies roughly as $E^{-2}$ as illustrated in Figure 3. This is due to the combined effect of a decreasing diffraction efficiency at the higher energies and the decreasing geometric area of the radial bands from which the higher energy photons can be diffracted and reach the detector. Laue lenses are peculiar optical systems because each specific photon energy is only collected from a limited radial band on the lens where Bragg's law is fulfilled. The outer radial bands correspond to the larger Bragg angles (lower energies) and the inner radial bands correspondingly collect only the higher energies.

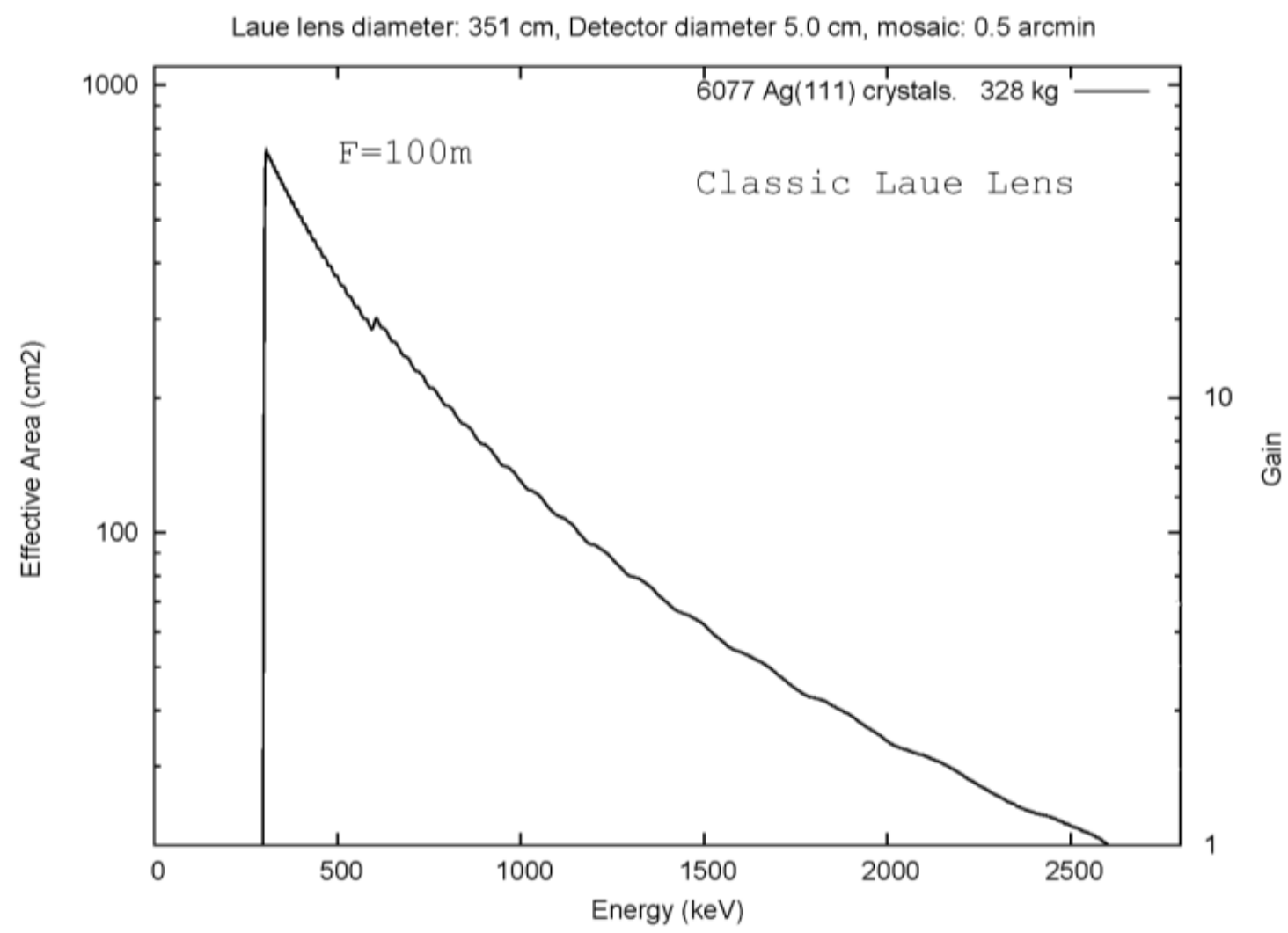

**Fig. 3** Effective area of a simple Laue-lens over an extended energy range.
The crystal thickness is optimized for peak reflectivity at 100 m focal length.
The 'Gain' (right hand scale) is the ratio between the lens effective area and the focal spot geometric area.
It expresses the factor by which the lens augments the target flux hitting the focal spot.
Total mass of crystals: 328 kg. Mass of individual crystals ranging from 36 to 475 g

Note that Bragg's law does not imply any true focusing of the radiation, each crystal simply diverts a beamlet with the same cross section as the physical crystal – the diverted beam maintains its shape as it travels along. Actually, this is only true for a strictly monochromatic beam. As the diffracted beam contains photons with some spread of energies, each photon is diffracted through its own proper Bragg angle and the diffracted beam will gradually spread out in a spectrum as it travels towards the focal plane. So although nominally our focal spot should be a 4 cm circle matching our crystals, we have allowed for some spectral defocusing when deciding the size of the detector. The spectral defocusing is included in all the performance calculations in the paper.

Errors in the alignment of the Bragg planes of the lens crystals will reduce the flux collected at the detector. We shall assume that the diffraction efficiency distribution of our crystals are Gaussian in shape with a full width at half maximum equal to the mosaic width, 30". Further, we assume that the alignment system, which we are developing, will be precise enough to ensure that the final alignment errors are negligible compared to the crystal mosaic width, i.e. the residual alignment errors shall be less than about 5".

It should also be emphasized that the 'Gain' indicated in the right hand ordinate scale on the figure is only a useful parameter when considering point sources, more specifically only for sources with an angular extent smaller that the mosaic width of the crystal. Unfortunately, this means that the strongest known celestial source of MeV-gamma rays, the 0.511 MeV source in the center of our own Galaxy may be quite inconspicuous as observed with a Laue lens. However, the centers of other, more distant galaxies may appear point like and should be very interesting targets.

We make a point of mentioning the total mass of crystals utilized in each of the discussed lens configuration because this is the dominant part in the total lens mass and past studies of Laue lenses for space have shown the weight of the lens to be a main factor driving the design.

## 5) Improving on the classic Laue lens

It is clear from Figure 3 that the classic Laue lens designed to cover the 511 keV band including the continuum below the peak is not suited for energies much above 1 MeV. We shall now discuss ways to improve the lens efficiency at the higher energies.

### 5.1) Energy tuning

If we can modify the tilt of each crystal, we may retune the lens for a different energy range (Note: this also implies a change in the focal length of the lens, but in a two-satellite configuration with the lens and the detector on separate spacecraft this is not a serious difficulty). In Figure 4 we compare the response of a single lens covering the full energy range from 300 keV to 2700 keV, with the response of a truncated lens covering only 300 to 800 keV, but retuned to a sequence of other focal lengths and corresponding higher and lower energies. Eliminating the inner rings of the classical lens saves significantly on the total crystal mass. Obviously, we hereby loose the possibility to observe all energies simultaneously, but this price is worth paying, considering the improved access to both higher and lower energies provided by the lens tuning.

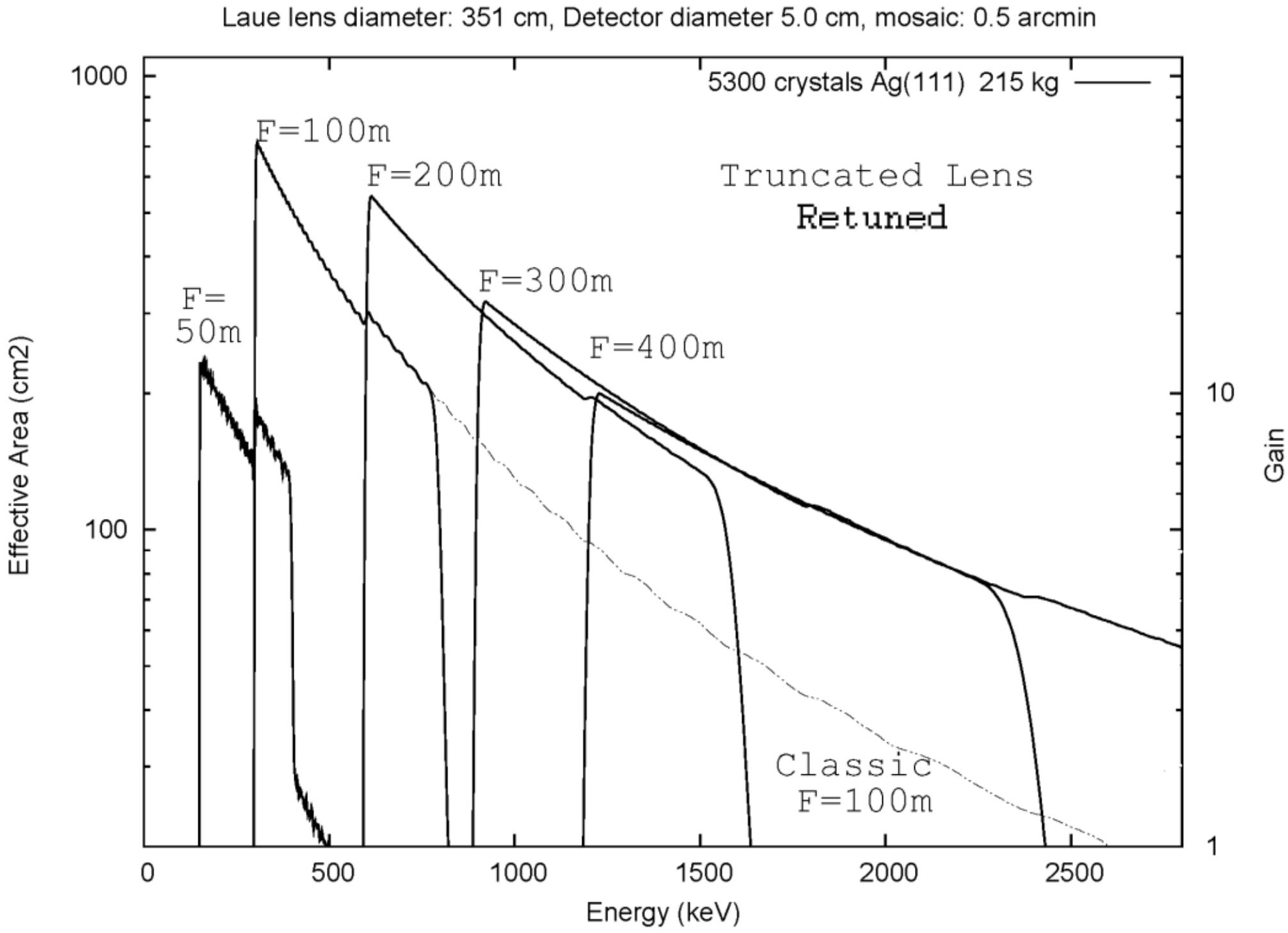

**Fig. 4** Effective area of single layer lens designed to cover from 300 to 800 keV at F=100 m, and sequentially retuned to other focal lengths. Total mass of crystals: 215 kg. Mass of individual crystals ranging from 36 to 149 g

In Figure 4 we note the significant increase of the lens effective area at high energies for increasing focal lengths and also a strong decrease of the lens effective area when the focal length is decreased below 100 m. This increase of the lens effective area with increasing focal length is discussed in more detail in the appendix. The magnitude of the effect is maybe surprising. One significant factor is the ratio between the angular diameter of the crystals as seen from the detector and the mosaic width of the crystals. This ratio will decrease for long focal lengths, thereby increasing the effective diffraction efficiency of the crystals. Conversely, for shorter focal lengths the average efficiency of the lens crystal will decrease significantly.

One may ask if we could not increase the lens efficiency at the shorter focal lengths by choosing a larger mosaic width for our crystals? Indeed we could, but this would require thicker crystals (larger mass) and actually lead to a reduction of the lens efficiency at the higher energies because an increase in the mosaic width will reduce the peak reflectivity of the crystals. Thus, a mosaic width of 0.5 arc minutes appear to be a good compromise.

### 5.2) Double crystal layers

The efficiency of a Laue-lens may be improved by about 65 % by stacking two crystal layers on top of each other. The peak diffraction efficiency for a single crystal layer is limited primarily by the competition between coherent diffraction and coherent back-diffraction of the photons. Incoherent loss processes play a lesser role. Therefore, we can stack two layers on top of each other with only limited mutual absorption losses, as long as the two layers do not coherently diffract the same photon energies. Figure 5 illustrates the improvement obtained by stacking two crystal layers, Ag(111) and Ag(200). The thickness which maximizes the total response in the double layer configuration is about 65 % of the optimal thickness for a single layer.

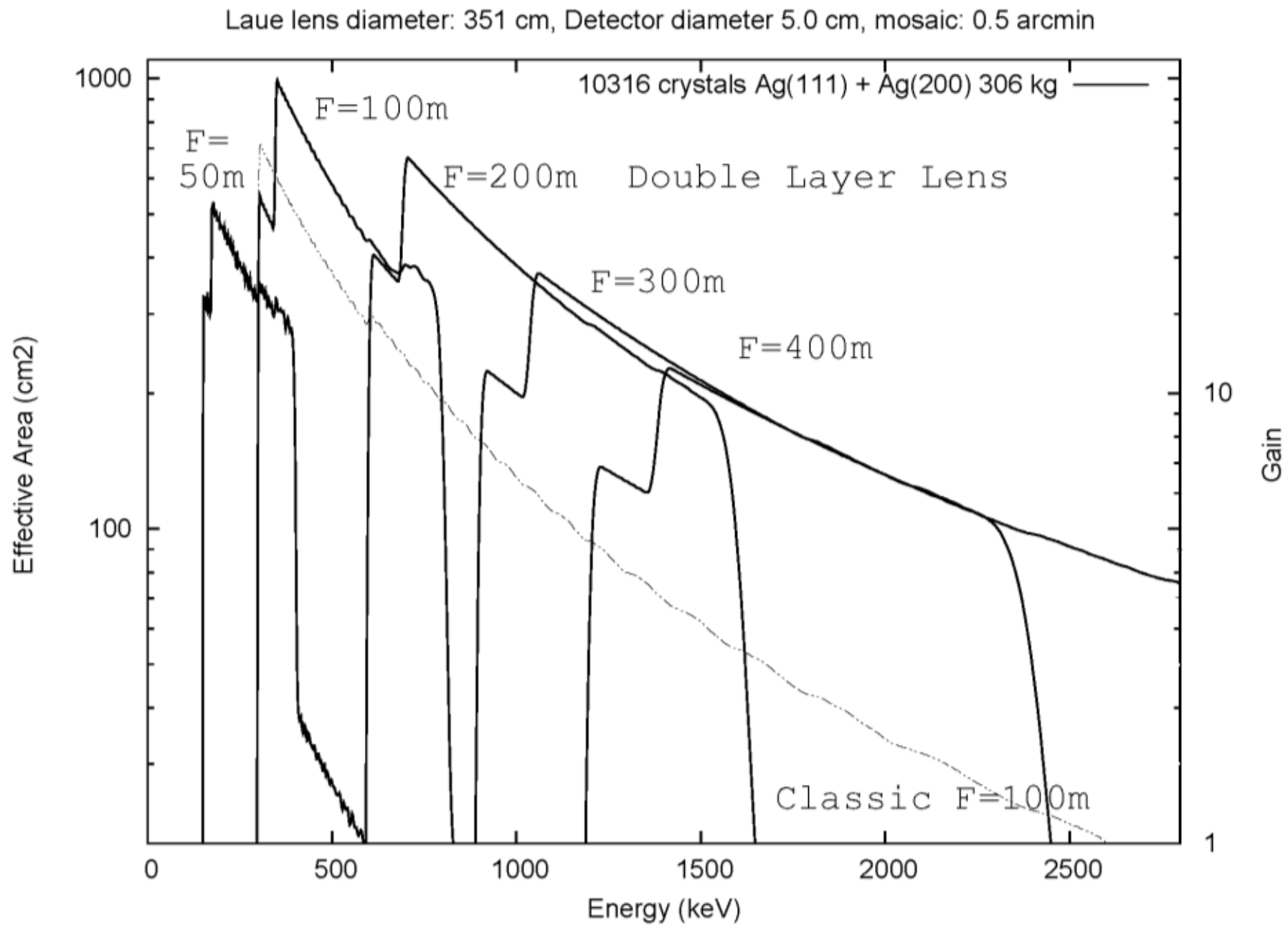

**Fig. 5** Double layer lens based on Ag(111) and Ag(200) crystals. The crystal thickness is chosen to optimize the response at F=100 m Total mass of crystals: 306 kg. Mass of individual crystals ranging from 23 to 104 g

It interesting to note that although we increased the total crystal mass relative to the single layer lens, the lens efficiency actually increased, most significantly for F=50 m, i.e. for the lowest energies. The important point here is that the average crystal thickness is reduced. This illustrates that the coherent back diffraction is a much stronger effect than the incoherent absorption.

In the accompanying technology paper (Lund, 2020) we discuss the practical feasibility of double crystal layers. Using more than two layers, unfortunately, does not provide significant further improvements.

### 5.3) Balancing the lens response

By replacing the basic crystal choice with other crystals or by employing diffraction from crystal planes corresponding to higher Miller-index values it is possible to shift part of the highly peaked response at lowest energies to higher energies. In Figure 6, we show one possible lens configuration with improved high energy response. Comparing to the photon collection power of the fixed-energy, one-layer lens we have now achieved a six-fold increase in the response at 2.4 MeV.

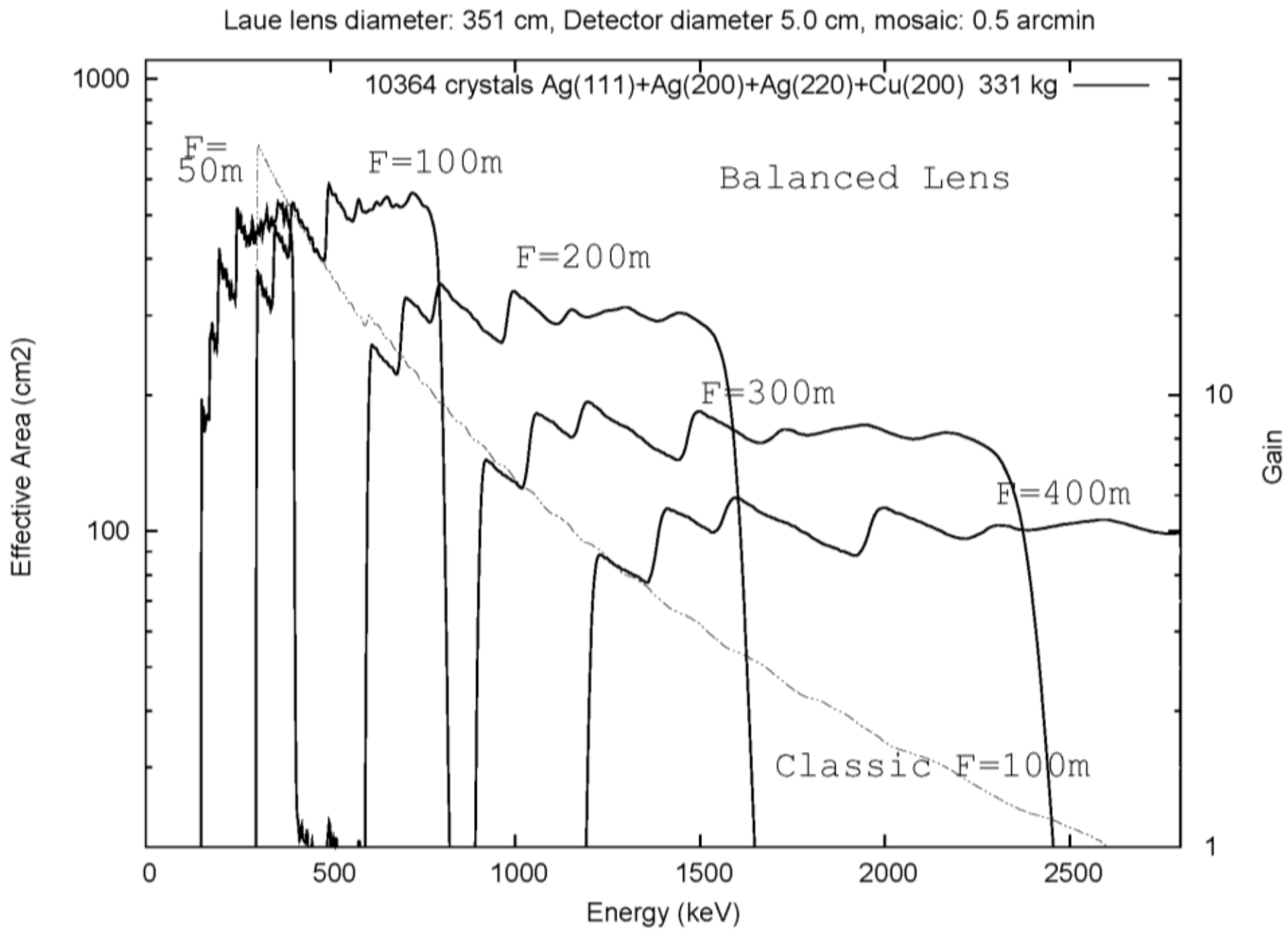

**Fig. 6** Example of a possible lens design with improved high energy response.
Total mass of crystals: 306 kg. Mass of individual crystals ranging from 23 to 104 g

## 6) Discussion

The Laue lens designs presented here are examples of what may be achieved once we master the technologies needed for adjusting the tilt angle of the individual crystals. Of course, the calculated efficiencies are idealized, assuming that all crystal parameters are well under control and mounting and alignment errors are small compared to the mosaic assumed 30" mosaic width.

To illustrate the power of such lenses we have used the Crab flux as measured by SPI on INTEGRAL (Jourdain and Roques 2009) and the lens effective areas from Figure 6. Our anticipated photon rates traversing a 5 cm diameter focal spot are shown in the table below.

| F (m) | Mean E (keV) | Lens $\Delta$E (keV) | SPI Crab flux (ph/cm$^2$skeV) | Balanced lens area (cm$^2$) | Balanced lens Crab rate (photons/s per $\Delta$E) |
|---|---|---|---|---|---|
| 50 | 275 | 250 | $7\times10^{-5}$ | 440 | 7.7 |
| 100 | 550 | 500 | $1.5\times10^{-5}$ | 490 | 3.7 |
| 200 | 1100 | 1000 | $4\times10^{-6}$ | 290 | 1.2 |
| 300 | 1650 | 1500 | $2\times10^{-6}$ | 160 | 0.5 |

Table 1. Expected photon rates at the detector from the Crab. Note that the Crab rates here are calculated based on the assumption that the high energy emission originates in a 'point like' region, i.e. with an angular extent less than the mosaic width of the crystals used in the lens. The X-ray images of the Crab nebulae obtained from the Chandra telescope suggests a 1' diameter of the emission source around 1 keV, we expect the angular extent of the source to be smaller at higher energies.

We expect that a detector based on modern Compton imaging technology (Chiu et al, 2015) will have an efficiency of 25% or better across all the energy bands relevant here and that the background rates per unit detector volume will be very much smaller than those observed with the SPI instrument. At this point we shall not venture further estimating the limiting flux for the telescope. Such estimates, will be needed at a later stage.

The absorption of the gamma rays in the lens panel has not been included in the calculations presented here. We assume that for our purpose the support panel can be realized as an aluminum honeycomb panel – with our alignment possibilities we can compensate for small thermal distortions during flight. The internal cell structure of the panel can be adapted to leave room for the forest of alignment pedestals protruding through the panel. With such a design, we believe that an effective thickness of 4 mm aluminum will be adequate, this will corresponding to an absorption loss of a 5 to 10 percent dependent on energy.

One major problem facing all Laue telescope projects is the lack of suitable celestial calibration sources of sufficient intensity. The specific demand for a point source aggravates the problem. Pre-flight calibration of a full scale Laue telescope is probably not realistic. It is therefore possible that we will need to plan for not only a two-satellite system with a lens spacecraft and detector spacecraft, but even a three-satellite system including a third calibration spacecraft with an X-ray generator on board.

## 7) Conclusions

We have demonstrated how a Laue lens optimized to study the important electron-positron annihilation radiation may be converted in-flight to a lens covering the nuclear line region between 0.5 and 2.5 MeV. We also reiterated the advantages of using double layer lenses, both increasing the photon collection area by about 65 %, and facilitating the balancing of the photon collection efficiency across energies.

Simulations combining already existing detector designs (Chiu et al, 2015) with a lens as described in this paper indicate that continuum sensitivities better than $10^{-7}$ photons/cm$^2$ may be obtained with observation times of a few days.

In the accompanying paper (Lund, 2020) we describe our technology development work aimed at making the idea of a tunable, double layer Laue-lens a realistic proposition in the not too distant future.

## Appendix: Effective Area dependence on telescope focal length

The flux collection from a Laue lens at a specific photon energy, $E$, depends on the distance from the lens plane to the detector – the telescope focal length, $F$. We can identify three effects to consider:

1). **The geometric area of the diffracting ring.** For any specific value of $E$ photons are only collected on to the detector from a ring-shaped area on the lens surface. This ring is simply the front face of the detector projected back to the lens through an angle $2\Theta_B$, where $\Theta_B$, is the Bragg angle at this energy. The geometric area is proportional to the radius of this ring, i.e. to $F$.

2). **Crystal peak diffraction efficiency.** In general, the diffraction efficiency is decreasing with increasing energy. This is somewhat compensated by the decreasing Compton absorption at higher energies. But as the focal length increases the diffracting ring for energy, $E$, shifts outward on the lens surface. The crystals located here are thinner than optimal for the new $E$. All in all, the peak diffraction efficiency is goes down almost as fast as the ring radius increases.

3). **Average diffraction efficiency of contributing crystals**. Consider, for the sake of the argument, a set of square crystal facets of dimensions $d$x$d$, set in a spiral configuration as illustrated below. The diffracted beams from the crystals illuminates a circular detector of diameter, $d$. (This is maybe not the optimal detector diameter, but this is not important for the following discussion). The crystals are assumed to be mosaic, with a Gaussian distribution of crystallite orientations resulting in a rocking curve with a standard deviation of $\omega$. We shall assume that all the crystals are oriented such that the plane corresponding to the mean crystallite orientation (the peak of the rocking curve) passes through a point on the telescope axis at a distance $2F$ from the lens plane, behind the detector. This assures that for all crystals, the image of the crystal is well centered on the detector <u>at the peak of the diffraction efficiency</u> (which in this configuration corresponds to a different energy for every crystal).

The spiral configuration is not the crystal lay-out we use, but the geometry effects are simpler to visualize this way, and the average efficiency issue remains the same whatever crystal configuration is used.

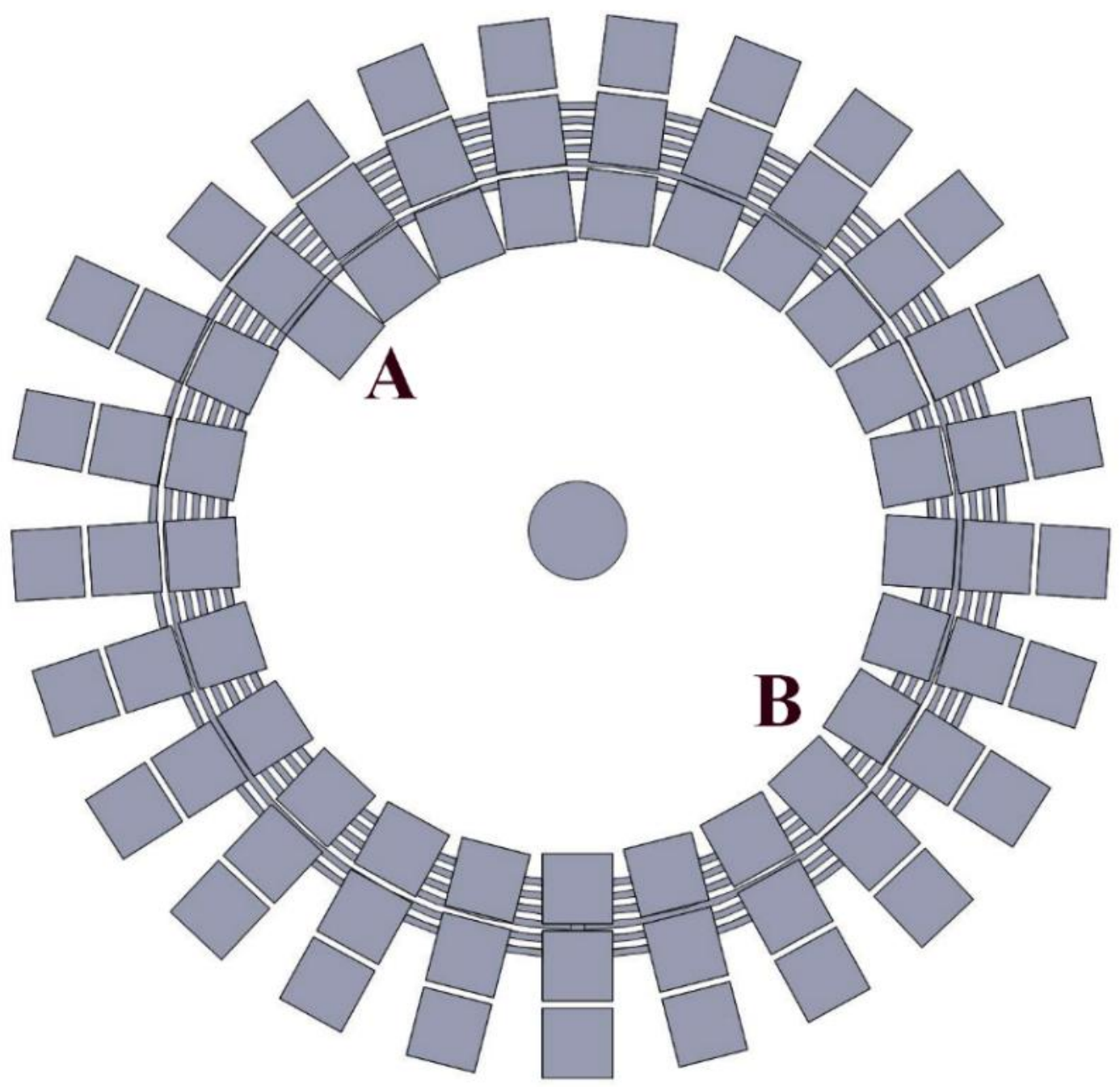

To guide the eye we have drawn a set of circles below the crystals indicating the geometric area from which photons of energy, *E*, can be diffracted and reach the detector.

The Laue lens is a very special optical system where photons of any particular energy can only be collected from such a specific ring-shaped area on the lens. Points outside this band may well diffract photons of energy, *E*, but the diffracted photons will miss the detector! Crystals near point A on the figure are ideally set for the energy, *E*, but those near point, B, are adjusted for slightly higher or lower energies – and consequently will not diffract energy, E, with their peak diffraction efficiency. Therefore, the effective diffraction efficiency must be averaged over the ring, taking into account the mosaic width of the crystals. The loss of diffraction efficiency becomes very noticeable at short focal lengths, where the angular width of the individual crystal facet in the lens as seen from the detector becomes comparable to the mosaic width.

The table below shows the calculated values of the focal length factor, the peak diffraction efficiency factor and the average efficiency factor relevant for the tunable lens discussed in the paper. Note that the average factor comes on top of the peak factor. The average only describes the average of the (assumed) Gaussian distribution of the reflectivities for diffraction off the peak of the distribution. From the numbers in the table it is clear that the combination of the focal length factor and the average efficiency factor significantly outweighs the drop in the peak efficiency factor for increasing focal lengths up to about 300 m.

| *F / E* | | 400 keV | | 700 keV | | 1400 keV | |
|---|---|---|---|---|---|---|---|
| Focal length | Focal length Factor | Peak efficiency factor | Average efficiency factor | Peak efficiency factor | Average efficiency factor | Peak efficiency factor | Average efficiency factor |
| 50m | 0.5 | 0.21 | 0.20 | | | | |
| 100m | 1.0 | 0.30 | 0.32 | 0.23 | 0.33 | 0.13 | 0.31 |
| 200m | 2.0 | | | 0.17 | 0.58 | 011 | 0.58 |
| 300m | 3.0 | | | | | 0.07 | 0.76 |
| 400m | 4.0 | | | | | 0.05 | 0.85 |